\documentclass[ALICE,manyauthors]{cernphprep}
\usepackage[comma,square,numbers,sort&compress]{natbib}
\usepackage{hyperref}
\usepackage{lineno}
\usepackage{xspace}
\usepackage[T1]{fontenc}
\usepackage{orcidlink} 

\begin{document}
%

\newcommand{\pp}           {pp\xspace}
\newcommand{\ppbar}        {\mbox{$\mathrm {p\overline{p}}$}\xspace}
\newcommand{\XeXe}         {\mbox{Xe--Xe}\xspace}
\newcommand{\PbPb}         {\mbox{Pb--Pb}\xspace}
\newcommand{\pA}           {\mbox{pA}\xspace}
\newcommand{\pPb}          {\mbox{p--Pb}\xspace}
\newcommand{\AuAu}         {\mbox{Au--Au}\xspace}
\newcommand{\dAu}          {\mbox{d--Au}\xspace}

\newcommand{\s}            {\ensuremath{\sqrt{s}}\xspace}
\newcommand{\snn}          {\ensuremath{\sqrt{s_{\mathrm{NN}}}}\xspace}
\newcommand{\pt}           {\ensuremath{p_{\rm T}}\xspace}
\newcommand{\meanpt}       {$\langle p_{\mathrm{T}}\rangle$\xspace}
\newcommand{\ycms}         {\ensuremath{y_{\rm CMS}}\xspace}
\newcommand{\ylab}         {\ensuremath{y_{\rm lab}}\xspace}
\newcommand{\etarange}[1]  {\mbox{$\left | \eta \right |~<~#1$}}
\newcommand{\yrange}[1]    {\mbox{$\left | y \right |~<~#1$}}
\newcommand{\dndy}         {\ensuremath{\mathrm{d}N_\mathrm{ch}/\mathrm{d}y}\xspace}
\newcommand{\dndeta}       {\ensuremath{\mathrm{d}N_\mathrm{ch}/\mathrm{d}\eta}\xspace}
\newcommand{\avdndeta}     {\ensuremath{\langle\dndeta\rangle}\xspace}
\newcommand{\dNdy}         {\ensuremath{\mathrm{d}N_\mathrm{ch}/\mathrm{d}y}\xspace}
\newcommand{\Npart}        {\ensuremath{N_\mathrm{part}}\xspace}
\newcommand{\Ncoll}        {\ensuremath{N_\mathrm{coll}}\xspace}
\newcommand{\dEdx}         {\ensuremath{\textrm{d}E/\textrm{d}x}\xspace}
\newcommand{\RpPb}         {\ensuremath{R_{\rm pPb}}\xspace}

\newcommand{\nineH}        {$\sqrt{s}~=~0.9$~Te\kern-.1emV\xspace}
\newcommand{\seven}        {$\sqrt{s}~=~7$~Te\kern-.1emV\xspace}
\newcommand{\twoH}         {$\sqrt{s}~=~0.2$~Te\kern-.1emV\xspace}
\newcommand{\twosevensix}  {$\sqrt{s}~=~2.76$~Te\kern-.1emV\xspace}
\newcommand{\five}         {$\sqrt{s}~=~5.02$~Te\kern-.1emV\xspace}
\newcommand{\twosevensixnn}{$\sqrt{s_{\mathrm{NN}}}~=~2.76$~Te\kern-.1emV\xspace}
\newcommand{\fivenn}       {$\sqrt{s_{\mathrm{NN}}}~=~5.02$~Te\kern-.1emV\xspace}
\newcommand{\LT}           {L{\'e}vy-Tsallis\xspace}
\newcommand{\GeVc}         {Ge\kern-.1emV/$c$\xspace}
\newcommand{\MeVc}         {Me\kern-.1emV/$c$\xspace}
\newcommand{\TeV}          {Te\kern-.1emV\xspace}
\newcommand{\GeV}          {Ge\kern-.1emV\xspace}
\newcommand{\MeV}          {Me\kern-.1emV\xspace}
\newcommand{\GeVmass}      {Ge\kern-.2emV/$c^2$\xspace}
\newcommand{\MeVmass}      {Me\kern-.2emV/$c^2$\xspace}
\newcommand{\lumi}         {\ensuremath{\mathcal{L}}\xspace}

\newcommand{\ITS}          {\rm{ITS}\xspace}
\newcommand{\TOF}          {\rm{TOF}\xspace}
\newcommand{\ZDC}          {\rm{ZDC}\xspace}
\newcommand{\ZDCs}         {\rm{ZDCs}\xspace}
\newcommand{\ZNA}          {\rm{ZNA}\xspace}
\newcommand{\ZNC}          {\rm{ZNC}\xspace}
\newcommand{\SPD}          {\rm{SPD}\xspace}
\newcommand{\SDD}          {\rm{SDD}\xspace}
\newcommand{\SSD}          {\rm{SSD}\xspace}
\newcommand{\TPC}          {\rm{TPC}\xspace}
\newcommand{\TRD}          {\rm{TRD}\xspace}
\newcommand{\VZERO}        {\rm{V0}\xspace}
\newcommand{\VZEROA}       {\rm{V0A}\xspace}
\newcommand{\VZEROC}       {\rm{V0C}\xspace}
\newcommand{\Vdecay} 	   {\ensuremath{V^{0}}\xspace}

\newcommand{\ee}           {\ensuremath{e^{+}e^{-}}} 
\newcommand{\pip}          {\ensuremath{\pi^{+}}\xspace}
\newcommand{\pim}          {\ensuremath{\pi^{-}}\xspace}
\newcommand{\kap}          {\ensuremath{\rm{K}^{+}}\xspace}
\newcommand{\kam}          {\ensuremath{\rm{K}^{-}}\xspace}
\newcommand{\pbar}         {\ensuremath{\rm\overline{p}}\xspace}
\newcommand{\kzero}        {\ensuremath{{\rm K}^{0}_{\rm{S}}}\xspace}
\newcommand{\lmb}          {\ensuremath{\Lambda}\xspace}
\newcommand{\almb}         {\ensuremath{\overline{\Lambda}}\xspace}
\newcommand{\Om}           {\ensuremath{\Omega^-}\xspace}
\newcommand{\Mo}           {\ensuremath{\overline{\Omega}^+}\xspace}
\newcommand{\X}            {\ensuremath{\Xi^-}\xspace}
\newcommand{\Ix}           {\ensuremath{\overline{\Xi}^+}\xspace}
\newcommand{\Xis}          {\ensuremath{\Xi^{\pm}}\xspace}
\newcommand{\Oms}          {\ensuremath{\Omega^{\pm}}\xspace}
\newcommand{\degree}       {\ensuremath{^{\rm o}}\xspace}

\begin{titlepage}
\PHyear{2026}       
\PHnumber{046}      
\PHdate{23 February}  

\title{$\pi$, K, and p production in high-multiplicity pp collisions at $\mathbf{\sqrt{\it s} = 13}$ TeV}
\ShortTitle{$\pi$, K, and p production in HM pp collisions at $\sqrt{s} = 13$ TeV}   

\Collaboration{ALICE Collaboration\thanks{See Appendix~\ref{app:collab} for the list of collaboration members}}
\ShortAuthor{ALICE Collaboration} 


\begin{abstract}
This paper presents the measurement of $\pi^{\pm}$, $\text{K}^{\pm}$, and p($\bar{\text{p}}$) production in high-multiplicity proton--proton collisions at $\sqrt{s}=13$ TeV at midrapidity ($|y| < 0.5$) using the ALICE detector at the LHC. The transverse-momentum ($\pt$) spectra of these particles are reported for three high-multiplicity classes. The results show a mass- and multiplicity-dependent hardening of the $\pt$ spectra and an enhancement of the p/$\pi$ ratio at intermediate $\pt$. These features are similar to those observed in heavy-ion collisions, where quark--gluon plasma formation is expected. The new measurements have extended the highest average charged-particle multiplicity density per unit of pseudorapidity achieved in pp collisions, roughly a factor five higher than that in average inelastic pp collisions, thereby reducing the multiplicity gap between small and large collision systems. In addition, the results are further compared with previously published measurements and with model calculations obtained using distinct tunes of the PYTHIA 8 Monte Carlo generator, as well as with predictions from the EPOS4. The comparison of the \pt-integrated K/$\pi$ and p/$\pi$ ratios across different collision systems and energies suggests that particle production scales with charged-particle multiplicity, rather than with collision energy or system size. While the PYTHIA 8 tunes and the EPOS4 model are able to reproduce some of these measurements, either quantitatively or qualitatively, none of them consistently describes all observed features of the data.

\end{abstract}
\end{titlepage}

\setcounter{page}{2} 


\section{Introduction} 
In high-energy nucleus--nucleus (AA) collisions, a hot and dense state of deconfined, strongly interacting matter, known as quark--gluon plasma (QGP), is formed~\cite{Busza_2018}. The ultrarelativistic energies available at the Large Hadron Collider (LHC) provide a unique opportunity to explore its properties. This system undergoes a hydrodynamic evolution~\cite{Heinz_2013,Bernhard_2016,Gardim_2020} and exhibits an enhanced production of (multi-)strange hadrons, long-range angular correlations, non-zero anisotropic flow harmonics, and suppression of high transverse momentum ($\pt$) particles in the dense medium as measured by the nuclear-modification factor~\cite{Koch:1986ud,PhysRevLett.116.132302,Adam_2016}. The light-flavor hadrons produced at low $\pt$, created through soft parton interactions, provide essential insights into the collective behavior in the QGP~\cite{Acharya_2020_2}.

The observation of a ``ridge-like'' structure in angular correlations between produced particles, as well as non-zero anisotropic flow coefficients, has also been reported in small systems such as proton--proton (pp) and proton--lead (p--Pb) collisions~\cite{Khachatryan_2017,ALICE:2021nir,Khachatryan:2010gv, Khachatryan:2016txc, Aad_2016mok, ALICE:2024pPbLongRange,flow_pPb}. Studies of high-multiplicity (HM) pp and p--Pb collisions have revealed several key features that resemble those observed in heavier collision systems, where the formation of the QGP is well established. These features include the observation of mass-dependent hardening of $\pt$ spectra, the increase of average transverse momentum ($\langle \pt \rangle$) with multiplicity, strangeness enhancement, and an enhanced baryon-to-meson ratio at intermediate $\pt$~\cite{2017,2016389,2013795,201329,PhysRevC.90.054901,Abelev_2014,Acharya_2019, Khachatryan:2017jnq, Acharya_2020}. Despite these similarities, the nuclear modification factor remains consistent with unity at high $\pt$. This contrasts with the suppression observed in heavy-ion collisions. The nuclear modification factor represents the ratio of particle yields measured in lead--lead (Pb--Pb) or p--Pb collisions to those in pp collisions scaled by the number of binary nucleon--nucleon collisions~\cite{Adam_2016_2,Abelev_2014_3}.

The study of hadron-to-pion ratios as a function of the average charged-particle multiplicity (\avdndeta) from pp to p--Pb and then to Pb--Pb system suggested that the particle production depends primarily on charged-particle multiplicity, independent of collision system or energy~\cite{2017,Acharya_2020}. This observation raises the question of whether there are common physical mechanisms in small and large systems.

To investigate the origin of collective-like effects in pp collisions, it is essential to use observables that are sensitive to these phenomena and to compare the results with theoretical model predictions. In particular, the transverse momentum spectra of identified particles provide important information, e.g., the mass-dependent spectral shape and mass ordering, which are commonly associated with radial flow~\cite{Acharya_2020}. Since collective phenomena in pp and p--Pb collisions can not be derived directly from first-principle Quantum Chromodynamics (QCD), their description often relies on phenomenological models~\cite{Sjostrand_2006,Sj_strand_2008,Corke_2011,Pierog:2015EPOSLHC,Werner:2010Hydro}. PYTHIA, a Monte Carlo (MC) event generator based on the Lund string model~\cite{Skands_2014}, is able to generate high-multiplicity pp events with color reconnection, enhanced strangeness production~\cite{2017}, radial-like expansion via color-rope fragmentation~\cite{Ortiz2017,Fischer_2017}, and anisotropies in the final-state momentum distributions arising from repulsive interactions among color strings (string shoving)~\cite{BIERLICH201858}. In addition, color-reconnection schemes beyond the leading-color (CLR-BLC) approximation introduce string junction topologies and provide a good description of the p/$\pi$ ratio at low $\pt$ in minimum bias (MB) pp collisions~\cite{ALICE:2023FlavorRatio}. The EPOS4 model implements a core-corona approach~\cite{Werner_2024_EPOS4_Concepts,Werner:2007}, in which strings in low-density regions fragment to form the corona, while the high-density core undergoes collective expansion. This approach is able to describe certain collective-like features, such as radial flow and strangeness enhancement, in small collision systems~\cite{Werner:2024EPOS4}. However, none of these theoretical approaches provides a consistent quantitative description of the collective-like effects in small collision systems~\cite{ALICE:2021nir,ALICE_2024_spherocityHMpp,Acharya_2020}. Small collision systems at the highest charged-particle multiplicities therefore offer essential input for constraining and improving existing theoretical models.

In this article, we present the measurements of primary charged particles, i.e. $\pi^{\pm}$, $\text{K}^{\pm}$, and p($\bar{\text{p}}$) at midrapidity ($|y| < $0.5) in pp collisions at $\sqrt{s}= 13$ TeV across three high-multiplicity classes using the ALICE detector~\cite{Collaboration_2008,ALICE:2022wpn}. PYTHIA 8 and EPOS4 are used as reference models for comparison with the experimental results. The new results extend the previous measurements of light-flavor hadron production from an average charged-particle multiplicity of 26.0 to 35.8, thus reducing the gap between pp and large collision systems. At such high multiplicities, pp events reach charged-particle densities that are comparable to those observed in peripheral (70--80\%) Pb--Pb collisions at $\sqrt{s_{\mathrm{NN}}} = 2.76~\text{TeV}$~\cite{Abelev_2013f}.
This paper is organized as follows: Section~\ref{sec:Exper_setup_data_analysis} describes the ALICE experimental setup and data analysis. The results, along with model predictions, are presented and discussed in Section~\ref{sec:Results_discussion}. Finally, the paper is summarized in Section~\ref{sec:Summary}.

\section{Experimental setup and data analysis}
\label{sec:Exper_setup_data_analysis}
\subsection{The ALICE detector}
A detailed description of the ALICE apparatus, as well as its performance, can be found in Refs.~\cite{Collaboration_2008, 2014}. The detector configuration used in this analysis corresponds to the Run 2 setup prior to the Long Shutdown 2 upgrade. The main detectors used in this analysis are the V0~\cite{Collaboration_2008}, the Inner Tracking System (ITS)~\cite{collaboration_2010}, the Time Projection Chamber (TPC)~\cite{Alme_2010}, and the Time-Of-Flight (TOF)~\cite{Carnesecchi_2019}. All of these detectors are located inside a 0.5 T solenoid magnetic field.

The V0 detector is used for triggering, multiplicity estimation, and background suppression. It consists of two scintillator arrays positioned along the beam axis on either side of the interaction point (IP), located at $z=340$ cm (V0A) and $z=-90$ cm (V0C). These arrays cover the pseudorapidity ranges $2.8<\eta<5.1$ and $-3.7<\eta<-1.7$ for V0A and V0C, respectively. The V0M multiplicity estimator obtained from the combination of V0A and V0C signals is used to classify events according to their measured multiplicity~\cite{Acharya_2019}.

The ITS is the tracking device closest to the interaction point. It is composed of six cylindrical layers of high-resolution silicon detectors, covering a pseudorapidity range of $|\eta| < 0.9$. The two innermost layers of ITS consist of hybrid Silicon Pixel Detectors (SPD), located at a radial distance of 3.9 cm and 7.6 cm from the beam axis, covering pseudorapidity ranges of $|\eta| < 2$ and $|\eta| < 1.4$, respectively. The outer four layers are composed of Silicon Strip (SSD) and Silicon Drift (SDD) Detectors, which are also used to identify particles by measuring the specific energy loss (d\textit{E}$/$d$x$) at low transverse momentum.

The TPC, together with the ITS~\cite{Alme_2010}, is the main tracking detector of the ALICE central barrel and covers the same pseudorapidity range as the ITS. It is a gaseous detector with radial and longitudinal dimensions of approximately 85 cm $<$ $r$ $<$ 247 cm  and $|z| < 250$ cm, respectively. The TPC identifies charged particles through the measurement of their specific energy loss and provides three-dimensional spatial information for each track, with up to 159 samples per track.

The TOF~\cite{Carnesecchi_2019} detector is based on the Multigap Resistive Plate Chamber (MRPC) technology. It covers the full azimuthal range for the pseudorapidity interval $|\eta| < 0.9$ and is cylindrically symmetric at an average distance of $\sim$4 m from the beam axis. The TOF detector identifies charged particles by measuring their time of flight. Since particles and antiparticles are produced in equal amounts at LHC energies~\cite{2013}, for the rest of this paper, we will use the notation $\pi$, $\text{K}$, and $\text{p}$ to denote ($\pi^{+}+\pi^{-}$), ($\text{K}^{+}+\text{K}^{-}$), and ($\text{p}+\bar{\text{p}}$), respectively.

\subsection{Data sample, event and track selection}
The results presented in this paper are obtained by combining two independent analyses, namely ITS stand-alone (ITSsa) and TPC--TOF. The analyzed data correspond to pp collisions at $\sqrt{s}=13~\mathrm{TeV}$ collected during the LHC Run~2 in 2016, with an integrated luminosity of $13~\mathrm{pb}^{-1}$ for the HM-triggered sample. After all event selections, approximately $8\times10^{7}$ and $8\times10^{6}$ events are used for the ITSsa and TPC--TOF analyses, respectively. High-multiplicity events are selected using the HM trigger, which is defined by a threshold on the relative V0M amplitude with respect to its average value $\text{V0M}/\langle\text{V0M}\rangle \gtrsim 4.9$, selecting the top 0.1$\%$ of the minimum bias events~\cite{2021d,2020}. Additionally, accepted events are required to have a primary vertex in the fiducial region $|z| <$ 10 cm. The events with out-of-bunch pileup were removed using timing information from the V0 counter. In-bunch pileup was identified using the SPD and removed from the analysis~\cite{2014}. The fraction of rejected events does not exceed about 3$\%$, and its impact on the measured spectra is negligible~\cite{ALICE:2022:LightNuclei13TeV}.

For the measurement of $\pi$, K, and $\text{p}$ spectra, we divided the high-multiplicity data into three classes: HM I, HM II, and HM III, corresponding to the top 0--0.01$\%$, 0.01--0.05$\%$, and 0.05--0.1$\%$ of the V0M/⟨V0M⟩ signal percentile, respectively. The corresponding average charged-particle multiplicity densities (\avdndeta), defined as the total number of charged particles produced per unit pseudorapidity in the \etarange{0.5} window~\cite{Abelev_2013}, are listed in Tab.~\ref{table:dNdEta}.

\begin{table}[!ht]
  \begin{center}
  \caption{Average charged-particle multiplicity density \avdndeta measured in the three event multiplicity classes used in this analysis.}
  \label{table:dNdEta}
    \begin{tabular}{|c|c|c|}
      \hline
            Multiplicity class & $\sigma /\sigma_{\mathrm{INEL}>0}$ ($\%$) & $\avdndeta_{|\eta| < 0.5}$ \\
      \hline
      HM I                & 0 -- 0.01                     & 35.8 $\pm$ 0.5                        \\
      HM II                & 0.01 -- 0.05                     & 32.2 $\pm$ 0.4                        \\
      HM III                & 0.05 -- 0.1                      & 30.1 $\pm$ 0.4                        \\
      \hline
    \end{tabular}
  \end{center}
\end{table}
The tracks were selected using track‐selection criteria similar to those described in Ref.~\cite{Acharya_2019}. Analyzed tracks are restricted to the pseudorapidity region $|\eta| < $ 0.8 to ensure good acceptance and performance. To ensure good track selection, we require at least 70 traversed TPC readout rows; the tracks are selected having goodness-of-fit, $\chi^{2}$ per TPC cluster (hits), to be less than 4. The track’s distance of closest approach (DCA) to the primary vertex along the longitudinal axis must satisfy $\text{DCA}_{z}< $ 2 cm, and in the transverse plane ($\text{DCA}_{xy}$), it is required to satisfy a $\pt$-dependent cut of less than 7 times the DCA resolution. ITSsa tracks are required to have at least four ITS clusters, including at least one in the SPD layers, and at least three in the SDD or SSD layers, with $\chi^{2}$ per ITS cluster required to be less than 2.5.

\subsection{Particle identification}
Particle identification (PID) is performed using the specific energy loss (d\textit{E}$/$d$x$) measured by the ITS and TPC, and the time-of-flight measured in the TOF. Low-$p_{\text{T}}$ particles (typically  0.1--0.75 GeV/$c$ depending on the particle species) are reconstructed using a dedicated tracking algorithm which only uses the ITS information. In the ITSsa, hadron identification is performed by calculating the energy loss measured in three or four outer ITS layers~\cite{2014}. The difference between the measured and expected value of d\textit{E}$/$d$x$ for each type of particle is calculated in multiples of standard deviations $\sigma$. The detailed description of this method is provided in Ref.~\cite{Acharya_2019}.

\begin{figure}[!ht]

  \centering
  {
    \includegraphics[width=0.7\textwidth]{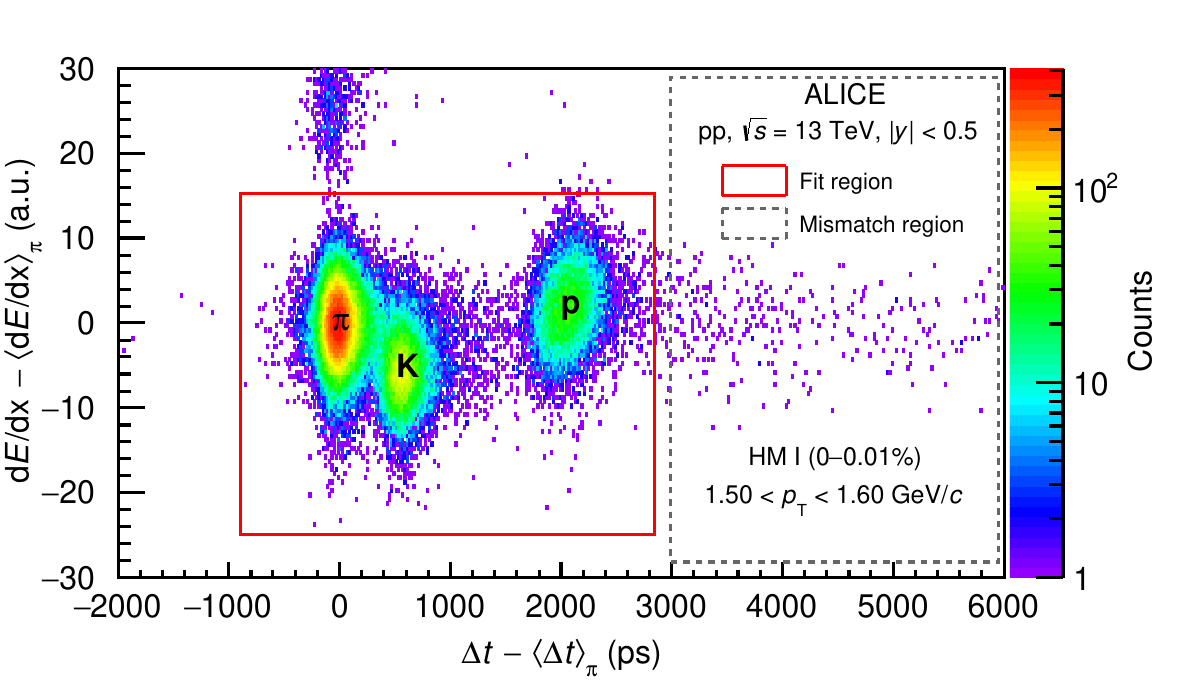}
  }
        \caption{Difference between measured ${\text{d}E}/{\text{d}x}$ (in the TPC) and its expected value for pions as a function of the same difference in time of flight as obtained from TOF. The plot is shown for 1.50 $< \pt <$ 1.60 GeV/$c$ at midrapidity. The red box represents the regions of the 2D fit, while the gray dashed box indicates the mismatch region.}
  \label{fig:performance}
\end{figure}
In the relatively higher $\pt$ region (0.6--4.0 GeV/$c$, or 0.7--4.0 GeV/$c$ depending on particle species), particle identification is performed using a TPC--TOF two-dimensional (2D) fit method. This method is illustrated in Fig.~\ref{fig:performance}. For each particle species
$i$ (pions, kaons, and protons), the $y$ axis shows $ X_{\text{i}} = {\text{d}E}/{\text{d}x}  -  { \langle {\text{d}E}/{\text{d}x} \rangle}_{\text{i}}$ obtained from the TPC~\cite{Acharya_2019} and the $x$ axis represents $ T_{\text{i}} =  \Delta t - {\langle \Delta t \rangle}_{\text{i}}$ provided by the TOF~\cite{2015d}. Here, ${\text{d}E}/{\text{d}x}$ and $\Delta$t are the measured specific energy loss in the TPC detector and time of flight in the TOF detector, respectively. The terms ${ \langle {\text{d}E}/{\text{d}x} \rangle}_{\text{i}}$ and ${\langle \Delta t \rangle}_{\text{i}}$ correspond to their expected values for a given $\pt$ range.

To extract the raw yield for each particle species in various $\pt$ intervals, a fit is performed on the $(T_i, X_i)$ distribution using the following fit functions.
\begin{equation}
        \frac{\text{d}^{2}N_{\text{tot}}}{\text{d}T_i\, \text{d}X_i} = \sum_j \frac{\text{d}^{2}N_j}{\text{d}T_i\, \text{d}X_i} + M_i,
\label{eq:rawYieldFit}
\end{equation}
where
\begin{equation}
\frac{\mathrm{d}^{2} N_j}{\mathrm{d}T_i \, \mathrm{d}X_i} \propto
\frac{\mathrm{d}N_j}{\mathrm{d}T_i} \times \frac{\mathrm{d}N_j}{\mathrm{d}X_i}
\end{equation}

and $M_i$ stands for the mismatch component that accounts for the background from wrongly associated TOF signals (TOF mismatches). Here, $i$ corresponds to the particle species under study, and $j$ runs over all the species (pions, kaons, and protons). The TOF mismatch is modeled using a data-driven template histogram constructed by randomly assigning TOF signals to reconstructed tracks~\cite{Abelev_2013f}. Figure~\ref{fig:performance} shows an example of $(T_i, X_i)$ distribution for $1.50 < p_{\mathrm{T}} < 1.60~\mathrm{GeV}/c$ under the pion mass assumption for the HM I multiplicity class. The various regions represent the different particle species, with pions located at $(0,0)$. The regions corresponding to other particles are separated due to the difference between their masses and the assumed pion-mass hypothesis. The red box represents the region of simultaneous fit for the TPC and TOF responses, while the gray dashed box indicates the mismatch region. The best fitting parameters are obtained by maximizing the logarithm of the likelihood of the fit function, and the raw yield for species~$i$ is extracted as the fit parameter.

\subsection{Corrections and Normalization}
The raw particle yields obtained from the ITSsa and TPC--TOF PID methods require various corrections, including tracking acceptance $\times$ efficiency, TPC--TOF matching efficiency, and contamination from secondary particles. The resulting spectra are then normalized by the number of events in each multiplicity class to obtain the final $\pt$-spectra~\cite{Acharya_2019,Acharya_2020,2021}. Tracking acceptance $\times$ efficiency accounts for the particles lost due to the geometrical acceptance of the detector as well as its limited detection efficiency. The TPC--TOF matching efficiency takes into account that not all analyzed tracks have a hit in the TOF. It is calculated as the ratio of the total number of tracks matched with the TOF to the total number of tracks reconstructed by the TPC. These corrections are derived from simulated pp collision events, which are generated using the PYTHIA 8 model (Monash 2013 tune~\cite{Sj_strand_2008,Skands_2014}), and the interactions of the generated particles with the experimental apparatus are modeled using GEANT4~\cite{AGOSTINELLI2003250}. In the simulation, one applies the same track selection criteria used for the data to account for detector acceptance and reconstruction efficiencies. It is also observed that these efficiencies show negligible variation with charged-particle multiplicity, so the multiplicity-integrated values are used for these corrections.

In this article, we present the measurements of primary particles, referred to as prompt particles in collisions, including decay products, except those from weak decays of strange particles and material interactions~\cite{ALICE-PUBLIC-2017-005}. The contamination from secondary particles produced in the detector material mainly affects protons, while that from weak decays affects both pions and protons. In order to estimate the contribution from secondary particles, produced from the weak decays or material interactions, a template fit method is used. This method is based on the fact that the DCA$_{xy}$ of primary particles, secondary particles from weak decays, and the secondaries from the material interactions are qualitatively different. Using the MC production, we obtain the corresponding DCA$_{xy}$ template distributions. To obtain the correct fraction of the primary particles, we fitted the data with the fit templates as carried out in the previous work~\cite{Acharya_2020}. The contribution of the secondary particles for pions and protons is maximum at low $\pt$ ($\sim$ 2.5 $\%$ at 0.1 GeV/$\it{c}$ for pions and 8.3$\%$ at 0.3 GeV/$\it{c}$ for protons) and becomes negligible at higher $\pt$, while the contribution from kaons is negligible.

\subsection{Systematic Uncertainties}
The systematic uncertainties in this analysis arise from various sources across the $\pt$ range. They are calculated by varying the selection criteria around the default values and comparing the resulting measurements with those obtained using the default criteria. A summary of these uncertainties is presented in Tab.~\ref{table:Systematics}.
\begin{table}[h]
\centering
	\caption{Sources of systematic uncertainties on the corrected spectra. The uncertainties are split into two categories, the common and the individual-analysis specific. The symbol ``--'' (negl.) indicates that the source is not applicable (negligible) for the corresponding analysis.}
\renewcommand{\arraystretch}{1.2}
\setlength{\tabcolsep}{5pt}
\label{table:Systematics}
\small

\begin{tabular}{l ccc c ccc c ccc}
\hline
& \multicolumn{11}{c}{\rule[-2ex]{0pt}{5ex}Uncertainty (\%)} \\
\cline{2-12}

& \multicolumn{3}{c}{\rule[-1.5ex]{0pt}{4ex}$\pi$} &
& \multicolumn{3}{c}{\rule[-1.5ex]{0pt}{4ex}K} &
& \multicolumn{3}{c}{\rule[-1.5ex]{0pt}{4ex}p} \\
\cline{2-4} \cline{6-8} \cline{10-12}

\rule[-1.5ex]{0pt}{4ex}
$p_\text{T}$ (GeV/$c$)
& 0.1 & 0.75 & 4.0
&
& 0.2 & 0.6 & 4.0
&
& 0.3 & 0.7 & 4.0 \\
\hline

& \multicolumn{11}{c}{\rule[-1.5ex]{0pt}{4ex}Common Source} \\
\hline
Correction for secondaries
& 1.0 & 1.0 & 1.0
&
& negl. & negl. & negl.
&
& 4.0 & 4.0 & 4.0 \\
Hadronic interactions
& 2.0 & 2.0 & 2.4
&
& 2.7 & 2.7 & 1.8
&
& 3.6 & 3.6 & 3.6 \\
\hline

& \multicolumn{11}{c}{\rule[-1.5ex]{0pt}{4ex}ITSsa} \\
\hline
PID
& 0.4 & 0.4 & --
&
& 3.3 & 3.3 & --
&
& 3.6 & 3.6 & -- \\
Material budget
& 4.8 & 0.3 & --
&
& 2.3 & 0.6 & --
&
& 5.0 & 0.9 & -- \\
$E \times B$ effect
& 3.0 & 3.0 & --
&
& 3.0 & 3.0 & --
&
& 3.0 & 3.0 & -- \\
Tracking efficiency (ITSsa tracks)
& 2.8 & 2.8 & --
&
& 2.9 & 2.9 & --
&
& 5.2 & 5.2 & -- \\
\hline

& \multicolumn{11}{c}{\rule[-1.5ex]{0pt}{4ex}TPC--TOF} \\
\hline
ITS-TPC matching efficiency
& -- & 0.7 & 1.5
&
& -- & -- & 1.5
&
& -- & -- & 1.5 \\
PID
& -- & 2.0 & 2.0
&
& -- & -- & 2.0
&
& -- & -- & 2.0 \\
TOF-matching efficiency
& -- & 3.0 & 3.0
&
& -- & -- & 6.0
&
& -- & -- & 4.0 \\
Material budget
& -- & 0.5 & 1.0
&
& -- & -- & 1.0
&
& -- & -- & 2.9 \\
Global tracking efficiency
& -- & 0.3 & 0.5
&
& -- & -- & 2.6
&
& -- & -- & 3.0 \\
\multicolumn{12}{l}{(incl.\ track cut variation)} \\
\hline
\rule[-1.5ex]{0pt}{4ex}
Total
& 7.1 & 6.7 & 5.3
&
& 6.7 & 6.4 & 7.6
&
& 10.4 & 9.3 & 8.5 \\
\hline
\end{tabular}
\end{table}
The final systematic uncertainty is obtained as the sum in quadrature of the uncertainties due to each variation. The sources of systematic uncertainties are divided into two parts: the first being common to ITSsa and TPC--TOF analyses, and the second specific to each analysis. The common sources of systematic uncertainty for both ITSsa and TPC--TOF analyses are the hadronic interaction cross section in the detector material and the corrections for secondary particles. The uncertainty associated with the hadronic interaction cross section in the detector material is evaluated using the GEANT4~\cite{AGOSTINELLI2003250} and FLUKA~\cite{Ferrari2005FLUKA} transport codes. Additionally, a 2\% systematic uncertainty is included to account for possible multiplicity dependence effects in track reconstruction efficiency and signal loss correction derived from MC simulations~\cite{Acharya_2020}. The contamination from out-of-bunch pileup tracks in the ITS--TPC matching is negligible, and the high-interaction rate related uncertainties are negligible relative to the other systematics.

The ITSsa analysis includes the uncertainties arising from particle identification (PID), the ITS material budget, and the Lorentz force ($E \times B$) effect. PID uncertainties are evaluated using a Bayesian approach and comparing the results obtained with the standard $n\sigma$ method~\cite{Adam_2016_2}. Material budget uncertainties are estimated by varying the ITS material by $\pm 7.5\%$~\cite{ALICE:2011pkp900}. The Lorentz force induces shifts in ITS cluster positions depending on the magnetic field polarity, resulting in an uncertainty of about  3\%. All of these uncertainties are taken from Ref.~\cite{Acharya_2020}. A detailed description of the estimation of systematic uncertainties for the ITSsa analysis can be found in Refs.~\cite{Acharya_2019,Acharya_2020}. The track selection uncertainties of ITSsa are obtained by varying the track selection criteria, mainly the $\text{DCA}_{xy}$ and $\text{DCA}_{z}$, the $\chi^2$ of the track and the number of ITS clusters.

The systematic uncertainties associated with TPC--TOF analysis include the uncertainties due to the ITS--TPC matching efficiency, PID, TOF-matching efficiency, material-budget, and tracking efficiency. The systematic uncertainties related to the ITS--TPC and TPC--TOF track matching are estimated by comparing the matching efficiencies in data and MC simulations. The material budget uncertainty is obtained by varying the material budget in the simulation by $\pm5\%$~\cite{Acharya_2020}. The uncertainties associated with tracking are assessed by varying the track selection criteria, which mainly include the number of clusters in the TPC, $\text{DCA}_{xy}$ and $\text{DCA}_{z}$. The uncertainties related to the particle identification are calculated by repeating the fits with varied fit ranges and parameters. The uncertainties listed in Tab.~\ref{table:Systematics} are assumed to be uncorrelated among different particle species, except those arising due to the magnetic field polarity, ITS--TPC matching efficiency, and the multiplicity-dependent tracking efficiency and signal loss corrections. The systematic uncertainties on the $\pt$-integrated observables will be described in Sec.~\ref{integrated yield} in detail.
\section{Results and discussion}
\label{sec:Results_discussion}
\subsection{\texorpdfstring{$\pt$-differential results}{pT-differential results}}
\begin{figure}[!ht]

  \centering
  {
    \includegraphics[width=0.9\textwidth]{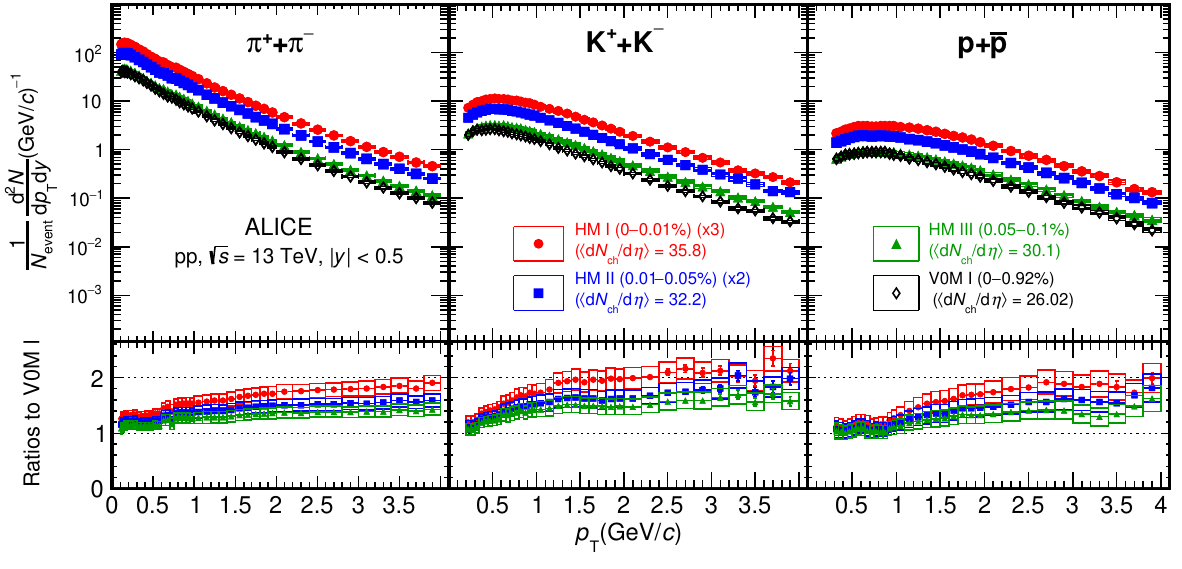}
  }
  \caption{(Top panels) Transverse momentum spectra of $\pi$, K, and $\text{p}$ for different multiplicity classes. Spectra are scaled to improve visibility. The corresponding ratios to V0M I (0--0.92$\%$) spectrum~\cite{Acharya_2020} are shown in the bottom panels. Vertical bars and open boxes represent the statistical and systematic uncertainties, respectively.}
  \label{fig:combinedresults}
\end{figure}
The $\pt$-differential spectra of the species under study ($\pi$, K, and p) for the multiplicity classes reported in Tab.~\ref{table:dNdEta} are shown in Fig.~\ref{fig:combinedresults}.
The ITSsa and TPC--TOF spectra overlap in the $\pt$ interval 0.6--0.75 GeV/${c}$ for pions. In this region, the spectra are combined using a weighted average based on the inverse squared uncorrelated systematic uncertainties~\cite{2015d}. For kaons and protons, however, there is no common $\pt$ interval between ITSsa and TPC--TOF. The exact $\pt$ ranges covered by ITSsa are 0.1--0.75 GeV/$c$, 0.2--0.6 GeV/$c$, and 0.3--0.7 GeV/$c$ for pions, kaons, and protons, respectively. For the TPC--TOF, these are 0.6--4.0 GeV/$c$ for pions and kaons and 0.7--4.0 GeV/$c$ for protons. The bottom panels of Fig.~\ref{fig:combinedresults} show the ratios of the spectra to the V0M I (0--0.92$\%$) multiplicity class measured by ALICE~\cite{Acharya_2020}. For all the identified hadrons, the ratios show a hardening of the spectra with increasing multiplicity. Such observations have been previously reported in pp, p--Pb, and Pb--Pb collisions and are attributed to hydrodynamical evolution of the system~\cite{Acharya_2020_2,Abelev_2014,Acharya_2020}.
\begin{figure}[!ht]

  \centering

  {
    \includegraphics[width=0.9\textwidth]{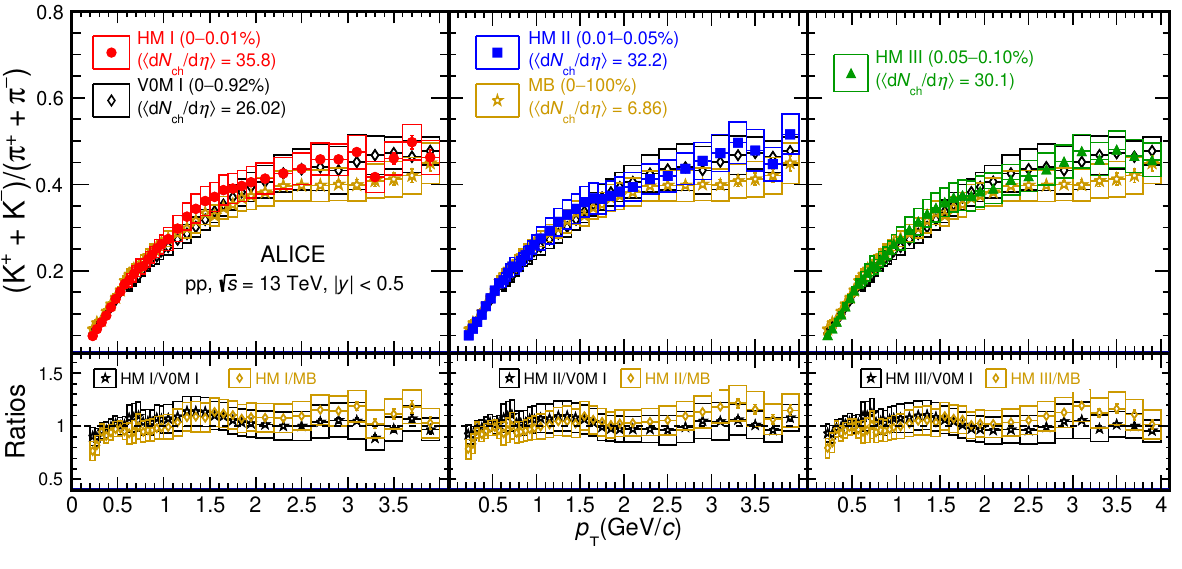}
  }
  \caption{(Top panels) K/$\pi$ spectrum ratios as a function of $p_{\text{T}}$ for all three high-multiplicity classes, V0M I (0--0.92$\%$)~\cite{Acharya_2020} and MB (0--100$\%$)~\cite{2021} multiplicity classes. The bottom panels show the ratio to the V0M I and MB classes. The boxes represent the systematic uncertainties, while the bars represent the statistical ones.}
  \label{fig:cobinedktopiratio23}
\end{figure}

The relative production of different charged hadrons at various transverse momenta can be studied through the ratio of their $\pt$-differential spectra as a function of event multiplicity. The measured K/$\pi$ ratio as a function of $\pt$ for different multiplicity classes is shown in Fig.~\ref{fig:cobinedktopiratio23}. 
\begin{figure}[!ht]

  \centering

  {
    \includegraphics[width=0.9\textwidth]{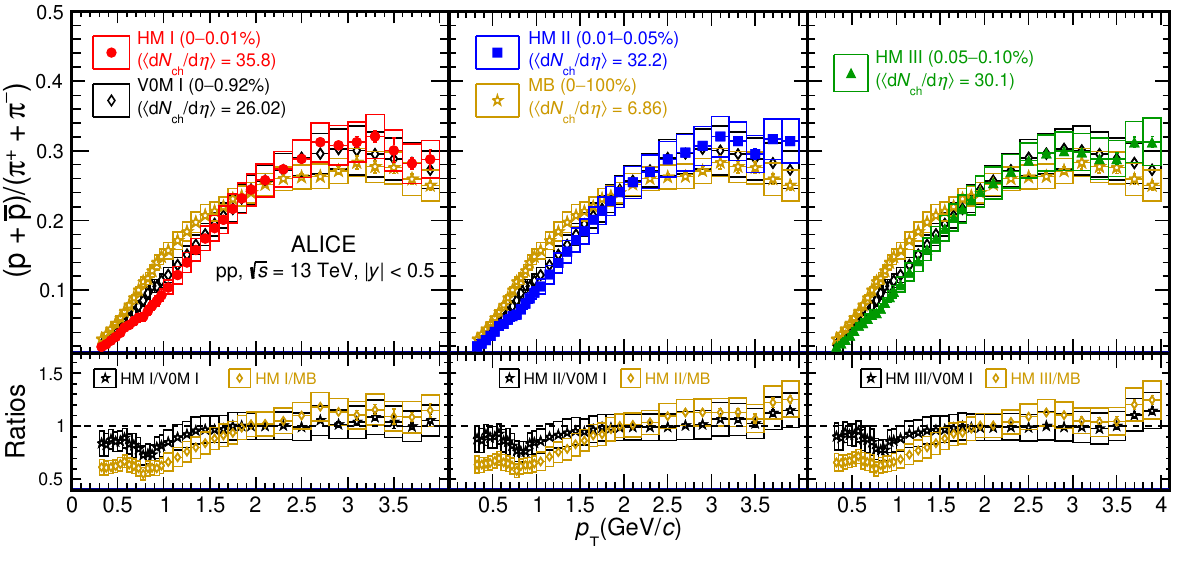}
  }
  \caption{(Top panels) p/$\pi$ spectrum ratios as a function of $p_{\text{T}}$ for three high-multiplicity classes and V0M I (0--0.92$\%$)~\cite{Acharya_2020} and minimum bias (MB) (0--100$\%$)~\cite{2021} multiplicity classes. The bottom panels show the ratio to the V0M I and MB classes. The boxes represent the systematic uncertainties, while the bars represent the statistical ones.}

  \label{fig:cobinedptopiratio23}
\end{figure}
The results are compared with those from the V0M I~\cite{Acharya_2020} and MB~\cite{2021} multiplicity classes measured in pp collisions at $\sqrt{s}=13$ TeV, as shown in Fig.~\ref{fig:cobinedktopiratio23}. The K/$\pi$ ratio exhibits an increasing trend at low $\pt$ and tends to saturate at higher $\pt$. The bottom panel shows the ratios of the measurements to V0M I and MB. The measurements are consistent with those of the V0M I and MB classes. Our measurements are also consistent with the previously published K/$\pi$ ratio in peripheral (70--80$\%$) Pb--Pb collisions at $\sqrt{s_{\text{NN}}}=2.76$~TeV and 5.02~TeV~\cite{Acharya_2020_2}, which shows a steep increase with $\pt$ from $\sim$0.07 at the lowest $\pt$ and saturates at $\sim$0.47 toward higher $\pt$. Figure~\ref{fig:cobinedptopiratio23} shows the p/$\pi$ ratio as a function of $\pt$ for different multiplicity classes. The results are also compared with the V0M I and MB multiplicity classes. The bottom panels show the ratio of the results to the published ones. A suppression in the p/$\pi$ ratio at low $\pt$ ($<$ 1 GeV/${c}$) and an enhancement at intermediate $\pt$  ($\sim$ 3 GeV/$c$) is observed. The results show a systematically increasing trend from HM I to HM III, although the significance of this rise is low due to large uncertainties. A similar enhancement of p/$\pi$ ratios at intermediate $\pt$ has been observed in p--Pb and Pb--Pb collisions, where it has been associated with collective flow or quark recombination~\cite{Abelev_2013f,Acharya_2020_2,Abelev_2014}. In addition, both the K/$\pi$ and p/$\pi$ ratios as a function of $\pt$ show a hardening of the spectra which is more pronounced for the heavier particles, see Fig.~\ref{fig:combinedresults}. Quantitatively, the K/$\pi$ (p/$\pi$) ratio increases by a factor of about 6 (15) from $\pt$ $\sim$ 0.3 to 3.0 GeV/c for the studied HM classes, suggesting a large increase for heavier particles (p). 
\subsection{\texorpdfstring{Comparison of $\pt$-differential results with models}{Comparison of pT-differential results with models}}
\begin{figure}[!ht]

  \centering

  {
    \includegraphics[width=0.9\textwidth]{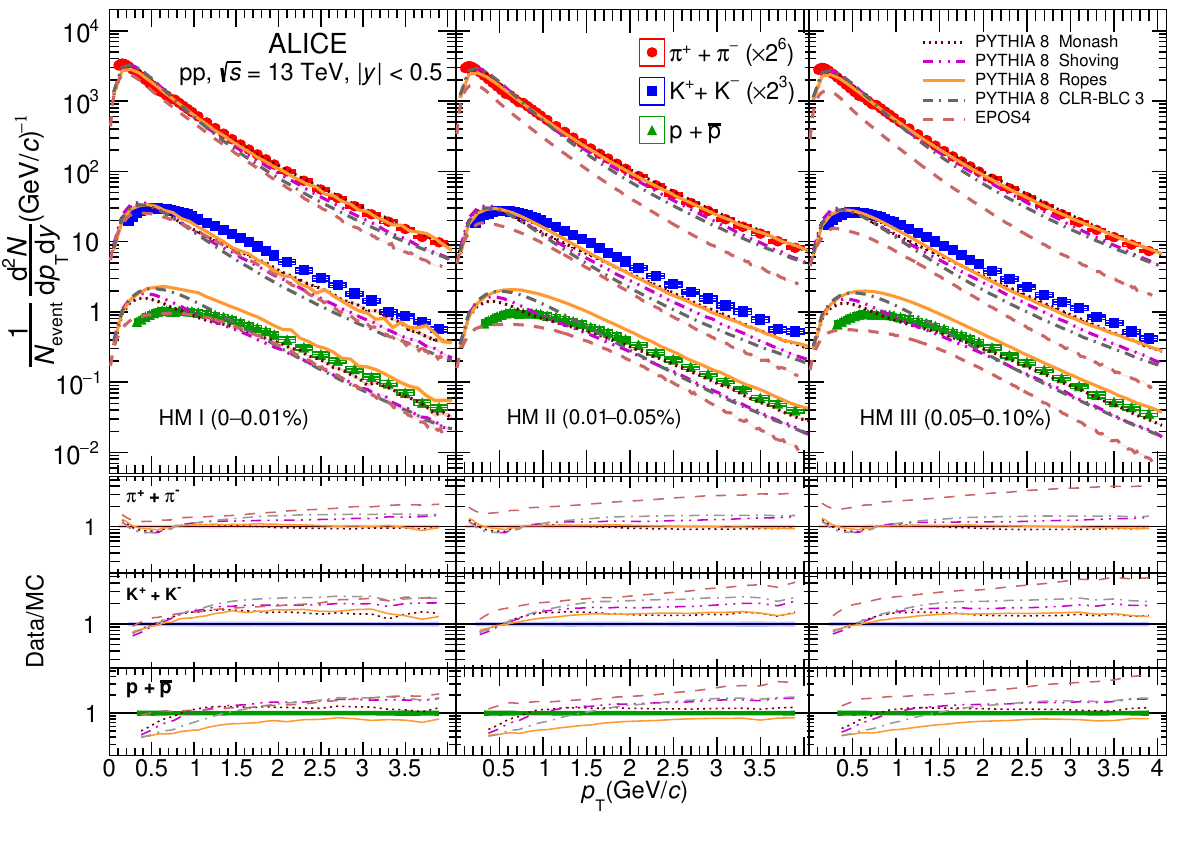}
  }

        \caption{(Top panels) Measured $\pt$ spectra of $\pi$, K, and p compared with PYTHIA 8 Monash, PYTHIA~8~Shoving, PYTHIA~8~Ropes, PYTHIA~8~CLR-BLC~3, and EPOS4 MC predictions for HM I (left), HM II (middle), and HM III (right) multiplicity classes. Statistical (bars) and systematic (open boxes) uncertainties are shown for the data. The shaded bands around the MC curves represent the statistical uncertainties of the model predictions. (Bottom panels) Ratio of the data to MC predictions. The bands around one in the bottom panels represent the systematic uncertainties of the data. Model uncertainties are not included in the ratio plots for better visibility.}
  \label{fig:SpectraModelComparison}
\end{figure}
The comparison between the measured $\pt$ spectra of $\pi$, K, and p with pQCD-inspired MC event generator calculations provides valuable insight into the hadronization mechanism. Figure~\ref{fig:SpectraModelComparison} presents a comparison of the measured $\pi$, K, and p spectra with predictions from different tunes of the PYTHIA 8 model, namely the Monash 2013 tune~\cite{Sj_strand_2008}, Ropes~\cite{bierlich2015effectsoverlappingstringspp}, Shoving~\cite{BIERLICH201858}, and CLR-BLC 3~\cite{Christiansen:2015yqa}, as well as with predictions from the EPOS4 model~\cite{Werner_2024_EPOS4_Concepts}, for different multiplicity classes.

PYTHIA 8 is based on the Lund String Model~\cite{Andersson:1983jt,Sjostrand:1984ic}, which describes hadronization through the fragmentation of QCD strings into hadrons. PYTHIA 8 with the Monash 2013 tune implements MPI and a basic version of the Color-Reconnection mechanism. PYTHIA 8 with ropes results in a higher effective string tension due to the overlap of several strings at the hadronization time (color ropes), leading to an increased production of baryons and strangeness~\cite{bierlich2015effectsoverlappingstringspp}. PYTHIA 8 with shoving tune accounts for string interactions, where strings repel each other in the transverse direction. This interaction generates microscopic transverse pressure, leading to long-range correlations~\cite{Kim_2021}. In PYTHIA 8, the CLR-BLC approximation implements the formation of string junction topologies during the color-reconnection stage, which subsequently leads to enhanced baryon production during string fragmentation. The \textsc{PYTHIA}\,8 CLR-BLC model provides several modes (0, 2, and 3) that impose different time-dilation and causality constraints on string reconnection before hadronization~\cite{Christiansen:2015yqa}. As discussed in recent studies, the Mode~3 provides a good description of the p/$\pi$ ratio at low $p_{\mathrm{T}}$ in MB pp collisions~\cite{ALICE:2023FlavorRatio}. Accordingly, the CLR-BLC~Mode~3 tune is compared with data and other models in high-multiplicity pp events. EPOS4, the latest version of the EPOS event generator, implements initial partonic scatterings using a Pomeron-based Gribov--Regge scheme and dynamically separates the system into a dense ‘core’ and a dilute ‘corona’~\cite{Werner_2024_EPOS4_Concepts,Werner:2007}. In these predictions, event classes are determined based on the number of charged particles simulated at both forward and backward pseudorapidities within the V0 detector coverage, using a method similar to that employed for data classification. Then, the $\avdndeta$ is determined in the central pseudorapidity region $|\eta| < $0.5.

\begin{figure}[!ht]

  \centering
  {
    \includegraphics[width=0.7\textwidth]{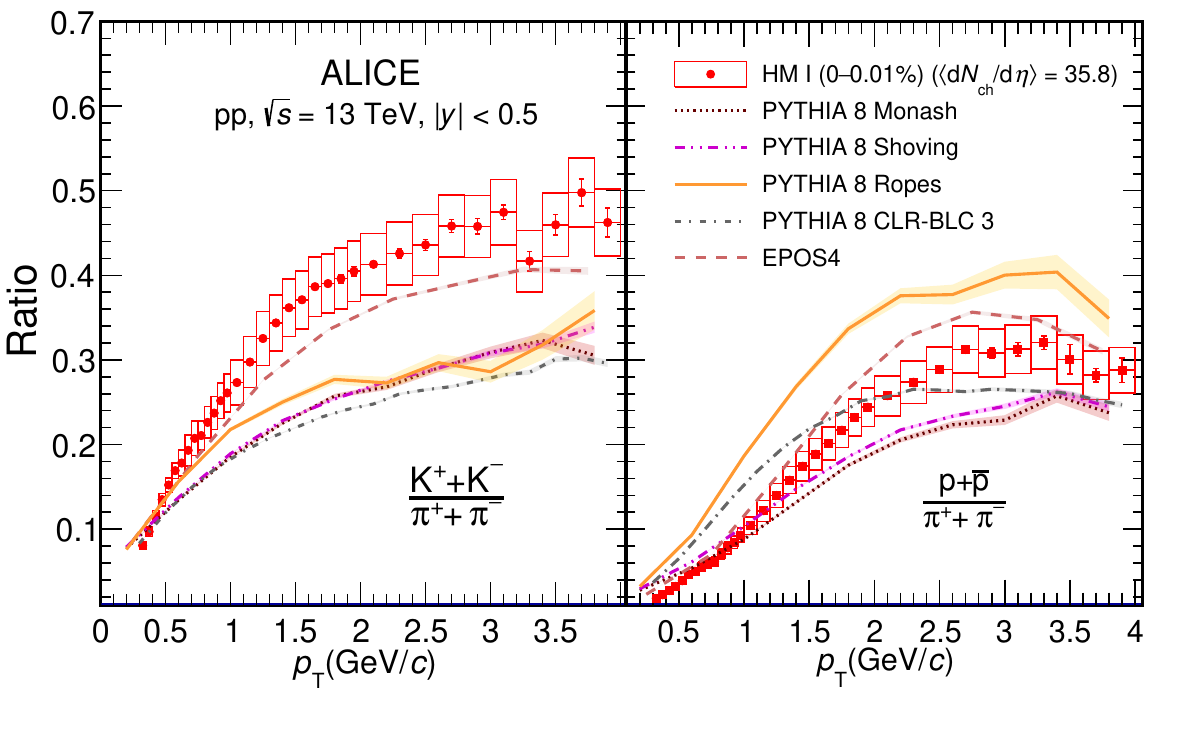}
  }

  \caption{($\text{K}^{+} + \text{K}^{-})$/$(\pi^{-} + \pi^{+}$) (left) and ($\text{p} + \bar{\text{p}})$/($\pi^{-} + \pi^{+}$) (right) ratios in pp collisions as a function of $p_{\text{T}}$ compared with PYTHIA 8 Monash, PYTHIA 8 Shoving, PYTHIA 8 Ropes, PYTHIA 8 CLR-BLC 3, and EPOS4 for the HM~I multiplicity class. The empty boxes are the systematic uncertainties of reported spectra ratios, and the bands represent the statistical uncertainties in the models.}
  \label{fig:ratioModel}
\end{figure}

It is observed from Fig.~\ref{fig:SpectraModelComparison} that the measured $\pt$ spectra of pions are qualitatively reproduced by the PYTHIA 8 Monash and PYTHIA 8 Ropes tunings within about 20\% over the full $\pt$ range, while the EPOS4 underestimates the spectra over the entire range. The comparisons of the K/$\pi$ and p/$\pi$ ratios, with PYTHIA 8 and EPOS4 model predictions are shown in Fig.~\ref{fig:ratioModel}.
All MC calculations underestimate the K/$\pi$ ratios with increasing momentum ($\pt>0.5$ GeV/$c$). Compared to the other PYTHIA 8 tunes, the CLR-BLC~3 shows an improved qualitative description of the p/$\pi$ ratios. EPOS4 predictions provide a better description of the data for both the K/$\pi$ and p/$\pi$ ratios compared to PYTHIA 8 tunes. It has also been observed that none of the PYTHIA 8 tunes capture the low-$\pt$ depletion seen in high-multiplicity results compared to the V0M I class.
\subsection{\texorpdfstring{$\pt$-integrated results}{pT-integrated results}}
\label{integrated yield}

The $\pt$-integrated particle yields ($\text{d}N/\text{d}y$) and mean transverse momentum ($\langle \pt \rangle$) are evaluated by integrating the measured $\pt$ spectra in the available $\pt$ ranges, while the L\'{e}vy--Tsallis functional fits are used to extrapolate the contribution to the  $\pt$ region that is not covered by the measurements~\cite{Tsallis1988}. The corresponding values of $\text{d}N/\text{d}y$ and $\langle \pt \rangle$ for the different multiplicity classes are reported in Tab.~\ref{Table:dndymeanpT}.

\begin{table}[!ht]
\centering
\caption{$\mathrm{d}N/\mathrm{d}y$ and $\langle p_{\mathrm{T}} \rangle$ (GeV/$c$) measured in different high-multiplicity classes. The first error represents the statistical uncertainty and the second one is the systematic.}
\renewcommand{\arraystretch}{1.15}
\setlength{\tabcolsep}{6pt}
\label{Table:dndymeanpT}
\small

\begin{tabular}{ccc}
\hline
\multicolumn{3}{c}{\rule[-1.8ex]{0pt}{4.5ex}$\pi^{+}+\pi^{-}$} \\
\hline
Multiplicity class & $\mathrm{d}N/\mathrm{d}y$ & $\langle p_{\mathrm{T}} \rangle$ (GeV/$c$) \\
\hline
HM I   & $33.967 \pm 0.011 \pm 1.915$ & $(6.234 \pm 0.004 \pm 0.160)\times10^{-1}$ \\
HM II  & $30.709 \pm 0.006 \pm 1.750$ & $(6.121 \pm 0.001 \pm 0.156)\times10^{-1}$ \\
HM III & $28.778 \pm 0.006 \pm 1.648$ & $(6.049 \pm 0.001 \pm 0.153)\times10^{-1}$ \\
\hline

\multicolumn{3}{c}{\rule[-1.8ex]{0pt}{4.5ex}$\text{K}^{+}+\text{K}^{-}$} \\
\hline
Multiplicity class & $\mathrm{d}N/\mathrm{d}y$ & $\langle p_{\mathrm{T}} \rangle$ (GeV/$c$) \\
\hline
HM I   & $5.067 \pm 0.007 \pm 0.380$ & $(10.862 \pm 0.012 \pm 0.274)\times10^{-1}$ \\
HM II  & $4.492 \pm 0.004 \pm 0.343$ & $(10.670 \pm 0.008 \pm 0.296)\times10^{-1}$ \\
HM III & $4.192 \pm 0.003 \pm 0.313$ & $(10.500 \pm 0.007 \pm 0.276)\times10^{-1}$ \\
\hline

\multicolumn{3}{c}{\rule[-1.8ex]{0pt}{4.5ex}$\text{p}+\bar{\text{p}}$} \\
\hline
Multiplicity class & $\mathrm{d}N/\mathrm{d}y$ & $\langle p_{\mathrm{T}} \rangle$ (GeV/$c$) \\
\hline
HM I   & $1.883 \pm 0.003 \pm 0.168$ & $(13.637 \pm 0.024 \pm 0.416)\times10^{-1}$ \\
HM II  & $1.696 \pm 0.002 \pm 0.151$ & $(13.455 \pm 0.017 \pm 0.410)\times10^{-1}$ \\
HM III & $1.590 \pm 0.002 \pm 0.142$ & $(13.163 \pm 0.017 \pm 0.396)\times10^{-1}$ \\
\hline
\end{tabular}
\end{table}

The fraction of the yields extrapolated with the functional fit are 9\% for $\pi$, 6\% for K, and 7\% for p. The order of these fractions remains approximately the same across all multiplicity classes. To estimate the systematic uncertainties of the extrapolated fractions, alternative fit functions, including Bose--Einstein, Fermi--Dirac, $\pt$ exponential, and Blast--Wave~\cite{2020} are used. The extrapolated systematic uncertainties in the integrated yield range from 0.9\% to 1.8\% for the $\pi$, K, and p particles, while for the $\langle p_\text{T} \rangle$, they vary from 0.83\% to 1.5\%. The statistical and systematic uncertainties on the $\text{d}N/\text{d}y$ and $\langle \pt \rangle$ are calculated by using the same methods as described in Ref.~\cite{Acharya_2019,Acharya_2020}.

\begin{figure}[!ht]

  \centering
  {
    \includegraphics[width=0.5\textwidth]{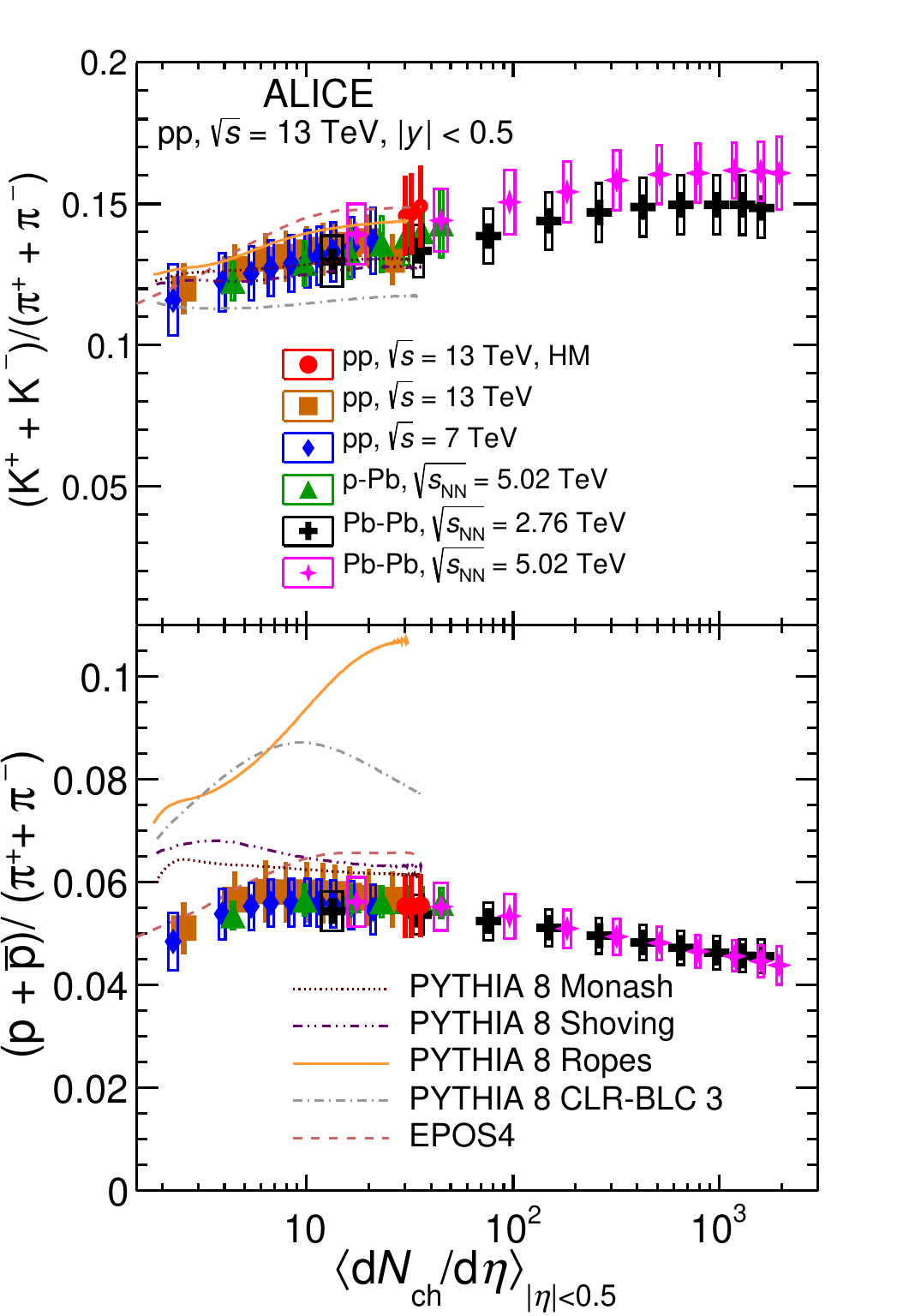}
  }
        \caption{K/$\pi$ (top) and p/$\pi$ (bottom) $\pt$-integrated ratios as a function of $\avdndeta$ measured in pp, p--Pb, and Pb--Pb collisions at different center-of-mass energies~\cite{Acharya_2020, Acharya_2019, Abelev_2014, Acharya_2020_2, Abelev_2013f}. Empty boxes represent the total systematic uncertainties and bars indicate the statistical ones. Colored lines represent the predictions from different tunes of PYTHIA 8 and EPOS4 for pp collisions at $\sqrt{s}=13$ TeV.}
  \label{fig:ktopiptopi}
\end{figure}

The $\pt$-integrated kaon-to-pion and proton-to-pion ratios measured in high-multiplicity pp collisions at $\sqrt{s}=13$ TeV are shown in Fig.~\ref{fig:ktopiptopi}. These results are also compared with those measured in pp, p--Pb, and Pb--Pb collisions at $\sqrt{s_{\text{NN}}}=7$, 13, 5.02, and 2.76 TeV, respectively~\cite{Acharya_2020, Acharya_2019, Abelev_2014, Acharya_2020_2, Abelev_2013f}.
The measurements show good agreement, within systematic uncertainties, with previously published results. These measurements have extended the K/$\pi$ and p/$\pi$ ratios in pp collisions to higher multiplicities. A small increase in the K/$\pi$ ratio by a factor of 1.39 $\pm$ 0.19 is seen when moving from low-multiplicity pp to the highest multiplicities in Pb--Pb collisions, hints at strangeness enhancement~\cite{Letessier2007}. Our results from high-multiplicity pp collisions follow this trend. The p/$\pi$ ratio decreases by a factor of 1.23 $\pm$ 0.15 from pp ($\avdndeta$ $\sim$ 3.9) to high-multiplicity Pb--Pb collisions, suggesting the antibaryon–baryon annihilation in heavy-ion collisions~\cite{Vovchenko_2022}. Our results for the p/$\pi$ ratio follow the trend exhibited here. Furthermore, the measurements are compared with the predictions from PYTHIA 8 and EPOS4. The PYTHIA 8 Monash, Shoving, and Ropes tunes describe the integrated K/$\pi$ ratio reasonably well (Fig.~\ref{fig:ktopiptopi}), although noticeable deviations are observed in the $p_{\mathrm{T}}$-differential ratios (see Fig.~\ref{fig:ratioModel}). 
In contrast, the PYTHIA~8~CLR-BLC~Mode~3 tune systematically underestimates the measured ratio, especially for $\langle \mathrm{d}N_{\mathrm{ch}}/\mathrm{d}\eta \rangle > 4$. It should be noted that both PYTHIA 8 Monash and Shoving provide a qualitative description of the p/$\pi$ ratio for $\avdndeta$ $>$ 5. However, the PYTHIA 8 Ropes and PYTHIA~8~CLR-BLC~3 overestimate the data. EPOS4 describes the K/$\pi$ ratio in the low ($\langle \mathrm{d}N_{\mathrm{ch}}/\mathrm{d}\eta \rangle < 6$) and high ($\langle \mathrm{d}N_{\mathrm{ch}}/\mathrm{d}\eta \rangle > 30$) multiplicity regions, while it reproduces the p/$\pi$ ratio up to $\avdndeta \sim 8$.
\begin{figure}[!ht]

  \centering
  {
    \includegraphics[width=0.9\textwidth]{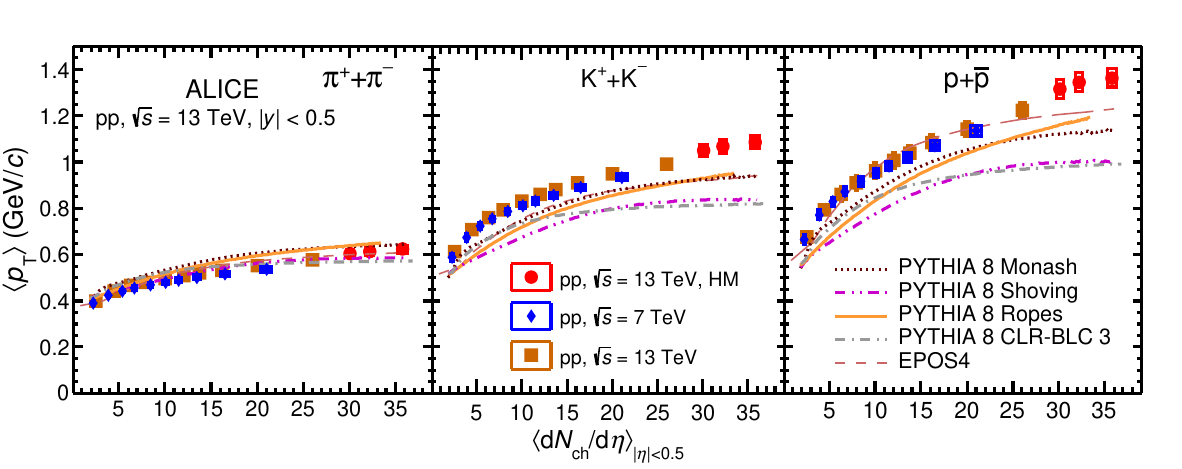}
  }
        \caption{$\langle \pt \rangle$ of $\pi$, K, and p as a function of $\avdndeta$ measured in pp collisions at $\sqrt{s}=7$ and 13 TeV~\cite{Acharya_2019, Acharya_2020}. Open boxes indicate systematic uncertainties, and bars indicate the statistical ones. Colored lines represent the predictions from different tunes of PYTHIA 8 and EPOS4 for pp collisions at $\sqrt{s}=13$ TeV.}
  \label{fig:meanpt}
\end{figure}

The $\langle \pt \rangle$ of $\pi$, K and p as a function of $\avdndeta$ along with the published results of pp collisions at $\sqrt{s}=7$ and 13 TeV are shown in Fig.~\ref{fig:meanpt}. A continuous increase in the $\langle \pt \rangle$ with multiplicity is observed, and this increase is steeper for the hadrons with higher mass. The new results confirm this trend. This observation is the same as observed in heavy-ion collisions, which is consistent with radial flow effects. Comparing pp and Pb--Pb at similar multiplicity suggests that $\langle \pt \rangle$ is higher in pp compared to that in Pb--Pb~\cite{Acharya_2019}. We also observe that all the models capture the increasing trend of $\langle \pt \rangle$ with $\langle \mathrm{d}N_{\mathrm{ch}}/\mathrm{d}\eta \rangle$. Among all the PYTHIA 8 tunes, PYTHIA 8 Shoving provides a quantitatively good description of the $\langle \pt \rangle$ of pions, whereas all the tunes underestimate the kaon and proton $\langle \pt \rangle$ values. The $\langle \pt \rangle$ of pions predicted by EPOS4 is consistent with the data over the full range of $\avdndeta$, whereas the $\langle \pt \rangle$ of protons is well reproduced by EPOS4 up to $\avdndeta$ $<$ 26, but is underestimated in the highest three multiplicity classes.
\section{Conclusion}
\label{sec:Summary}
In this article, results related to the production of $\pi$, K, and p in high-multiplicity pp collisions at $\sqrt{s}=13$~TeV are presented. These new measurements have extended previous studies of pp collisions by pushing the average charged-particle multiplicity densities ($\avdndeta$) from 26.0 to 35.8, further bridging the gap between small and large collision systems. The transverse-momentum spectra, integrated yields, mean transverse momentum, and K/$\pi$ and p/$\pi$ ratios are presented. A comparison of these results with the published data for different system sizes and energies, and with the PYTHIA 8 (Monash, Shoving, Ropes and CLR-BLC~3 tunes) and EPOS4 models is presented.

A hardening of the $p_{\mathrm{T}}$ spectra across multiplicity classes is observed, which is more pronounced for particles with larger masses. The $p_{\mathrm{T}}$-differential p/$\pi$ ratio shows an enhancement at intermediate $p_{\mathrm{T}}$, and the multiplicity-dependent K/$\pi$ and p/$\pi$ ratios display a smooth transition across different collision system sizes and energies. These results indicate that particle production is mainly driven by multiplicity, rather than by collision energy or system size. Furthermore, the $\langle p_{\mathrm{T}} \rangle$ as a function of $\langle \mathrm{d}N_{\mathrm{ch}}/\mathrm{d}\eta \rangle$ exhibits an increasing trend, with a more pronounced rise for heavier particles. These features, namely the hardening of $p_{\mathrm{T}}$ spectra with multiplicity and particle mass, the enhancement of the p/$\pi$ ratio at intermediate $p_{\mathrm{T}}$, and the increase of $\langle p_{\mathrm{T}} \rangle$, are similar to those observed in heavy-ion collisions, where they are associated with the presence of radial flow. The PYTHIA 8 Monash and Ropes tunes qualitatively reproduce the pion spectra, while the Monash and Shoving tunes provide a reasonable description of the integrated K/$\pi$ ratio. EPOS4 describes the increasing trend of the pion $\langle p_{\mathrm{T}} \rangle$ with multiplicity. Overall, EPOS4 captures the qualitative trends of the $p_{\mathrm{T}}$-dependent K/$\pi$ and p/$\pi$ ratios more consistently than the PYTHIA 8 tunes.

\newenvironment{acknowledgement}{\relax}{\relax}
\begin{acknowledgement}
\section*{Acknowledgements}

The ALICE Collaboration would like to thank all its engineers and technicians for their invaluable contributions to the construction of the experiment and the CERN accelerator teams for the outstanding performance of the LHC complex.
The ALICE Collaboration gratefully acknowledges the resources and support provided by all Grid centres and the Worldwide LHC Computing Grid (WLCG) collaboration.
The ALICE Collaboration acknowledges the following funding agencies for their support in building and running the ALICE detector:
A. I. Alikhanyan National Science Laboratory (Yerevan Physics Institute) Foundation (ANSL), State Committee of Science and World Federation of Scientists (WFS), Armenia;
Austrian Academy of Sciences, Austrian Science Fund (FWF): [M 2467-N36] and Nationalstiftung f\"{u}r Forschung, Technologie und Entwicklung, Austria;
Ministry of Communications and High Technologies, National Nuclear Research Center, Azerbaijan;
Rede Nacional de Física de Altas Energias (Renafae), Financiadora de Estudos e Projetos (Finep), Funda\c{c}\~{a}o de Amparo \`{a} Pesquisa do Estado de S\~{a}o Paulo (FAPESP) and The Sao Paulo Research Foundation  (FAPESP), Brazil;
Bulgarian Ministry of Education and Science, within the National Roadmap for Research Infrastructures 2020-2027 (object CERN), Bulgaria;
Ministry of Education of China (MOEC) , Ministry of Science \& Technology of China (MSTC) and National Natural Science Foundation of China (NSFC), China;
Ministry of Science and Education and Croatian Science Foundation, Croatia;
Centro de Aplicaciones Tecnol\'{o}gicas y Desarrollo Nuclear (CEADEN), Cubaenerg\'{\i}a, Cuba;
Ministry of Education, Youth and Sports of the Czech Republic, Czech Republic;
The Danish Council for Independent Research | Natural Sciences, the VILLUM FONDEN and Danish National Research Foundation (DNRF), Denmark;
Helsinki Institute of Physics (HIP), Finland;
Commissariat \`{a} l'Energie Atomique (CEA) and Institut National de Physique Nucl\'{e}aire et de Physique des Particules (IN2P3) and Centre National de la Recherche Scientifique (CNRS), France;
Bundesministerium f\"{u}r Forschung, Technologie und Raumfahrt (BMFTR) and GSI Helmholtzzentrum f\"{u}r Schwerionenforschung GmbH, Germany;
National Research, Development and Innovation Office, Hungary;
Department of Atomic Energy Government of India (DAE), Department of Science and Technology, Government of India (DST), University Grants Commission, Government of India (UGC) and Council of Scientific and Industrial Research (CSIR), India;
National Research and Innovation Agency - BRIN, Indonesia;
Istituto Nazionale di Fisica Nucleare (INFN), Italy;
Japanese Ministry of Education, Culture, Sports, Science and Technology (MEXT) and Japan Society for the Promotion of Science (JSPS) KAKENHI, Japan;
Consejo Nacional de Ciencia (CONACYT) y Tecnolog\'{i}a, through Fondo de Cooperaci\'{o}n Internacional en Ciencia y Tecnolog\'{i}a (FONCICYT) and Direcci\'{o}n General de Asuntos del Personal Academico (DGAPA), Mexico;
Nederlandse Organisatie voor Wetenschappelijk Onderzoek (NWO), Netherlands;
The Research Council of Norway, Norway;
Pontificia Universidad Cat\'{o}lica del Per\'{u}, Peru;
Ministry of Science and Higher Education, National Science Centre and WUT ID-UB, Poland;
Korea Institute of Science and Technology Information and National Research Foundation of Korea (NRF), Republic of Korea;
Ministry of Education and Scientific Research, Institute of Atomic Physics, Ministry of Research and Innovation and Institute of Atomic Physics and Universitatea Nationala de Stiinta si Tehnologie Politehnica Bucuresti, Romania;
Ministerstvo skolstva, vyskumu, vyvoja a mladeze SR, Slovakia;
National Research Foundation of South Africa, South Africa;
Swedish Research Council (VR) and Knut \& Alice Wallenberg Foundation (KAW), Sweden;
European Organization for Nuclear Research, Switzerland;
Suranaree University of Technology (SUT), National Science and Technology Development Agency (NSTDA) and National Science, Research and Innovation Fund (NSRF via PMU-B B05F650021), Thailand;
Turkish Energy, Nuclear and Mineral Research Agency (TENMAK), Turkey;
National Academy of  Sciences of Ukraine, Ukraine;
Science and Technology Facilities Council (STFC), United Kingdom;
National Science Foundation of the United States of America (NSF) and United States Department of Energy, Office of Nuclear Physics (DOE NP), United States of America.
In addition, individual groups or members have received support from:
Czech Science Foundation (grant no. 23-07499S), Czech Republic;
FORTE project, reg.\ no.\ CZ.02.01.01/00/22\_008/0004632, Czech Republic, co-funded by the European Union, Czech Republic;
European Research Council (grant nos. 101220549, 950692), European Union;
Deutsche Forschungs Gemeinschaft (DFG, German Research Foundation) ``Neutrinos and Dark Matter in Astro- and Particle Physics'' (grant no. SFB 1258), Germany;
CONVECS project, CUP C97H23001700002 FESR 2021-2027 program, Italy;
FAIR - Future Artificial Intelligence Research, funded by the NextGenerationEU program (Italy).

\end{acknowledgement}

\bibliographystyle{utphys}   
\bibliography{bibliography}

\newpage
\appendix

%
%

\section{The ALICE Collaboration}
\label{app:collab}
\begin{flushleft} 
\small

D.A.H.~Abdallah\,\orcidlink{0000-0003-4768-2718}\,$^{\rm 134}$, 
I.J.~Abualrob\,\orcidlink{0009-0005-3519-5631}\,$^{\rm 112}$, 
S.~Acharya\,\orcidlink{0000-0002-9213-5329}\,$^{\rm 49}$, 
G.~Aglieri Rinella\,\orcidlink{0000-0002-9611-3696}\,$^{\rm 32}$, 
L.~Aglietta\,\orcidlink{0009-0003-0763-6802}\,$^{\rm 24}$, 
N.~Agrawal\,\orcidlink{0000-0003-0348-9836}\,$^{\rm 25}$, 
Z.~Ahammed\,\orcidlink{0000-0001-5241-7412}\,$^{\rm 132}$, 
S.~Ahmad\,\orcidlink{0000-0003-0497-5705}\,$^{\rm 15}$, 
I.~Ahuja\,\orcidlink{0000-0002-4417-1392}\,$^{\rm 36}$, 
Z.~Akbar$^{\rm 79}$
% ...

\section*{Affiliation Notes}

$^{\rm I}$ Deceased\\
$^{\rm II}$ Also at: Fondazione Bruno Kessler (FBK), Trento, Italy\\
$^{\rm III}$ Also at: Max-Planck-Institut fur Physik, Munich, Germany\\
$^{\rm IV}$ Also at: Czech Technical University in Prague, Prague, Czech Republic\\
$^{\rm V}$ Also at: Instituto de Fisica da Universidade de Sao Paulo\\
$^{\rm VI}$ Also at: Dipartimento DET del Politecnico di Torino, Turin, Italy\\
$^{\rm VII}$ Also at: University College of Dublin, Dublin, Ireland\\
$^{\rm VIII}$ Also at: Department of Applied Physics, Aligarh Muslim University, Aligarh, India\\
$^{\rm IX}$ Also at: Institute of Theoretical Physics, University of Wroclaw, Wroclaw, Poland\\
$^{\rm X}$ Also at: Facultad de Ciencias, Universidad Nacional Aut\'{o}noma de M\'{e}xico, Mexico City, Mexico\\

\section*{Collaboration Institutes}

$^{1}$ A.I. Alikhanyan National Science Laboratory (Yerevan Physics Institute) Foundation, Yerevan, Armenia\\
$^{2}$ AGH University of Krakow, Cracow, Poland\\
$^{3}$ Bogolyubov Institute for Theoretical Physics, National Academy of Sciences of Ukraine, Kyiv, Ukraine\\
$^{4}$ Bose Institute, Department of Physics  and Centre for Astroparticle Physics and Space Science (CAPSS), Kolkata, India\\
$^{5}$ California Polytechnic State University, San Luis Obispo, California, United States\\
$^{6}$ Central China Normal University, Wuhan, China\\
$^{7}$ Centro de Aplicaciones Tecnol\'{o}gicas y Desarrollo Nuclear (CEADEN), Havana, Cuba\\
$^{8}$ Centro de Investigaci\'{o}n y de Estudios Avanzados (CINVESTAV), Mexico City and M\'{e}rida, Mexico\\
$^{9}$ Chicago State University, Chicago, Illinois, United States\\
$^{10}$ China Nuclear Data Center, China Institute of Atomic Energy, Beijing, China\\
$^{11}$ China University of Geosciences, Wuhan, China\\
$^{12}$ Chungbuk National University, Cheongju, Republic of Korea\\
$^{13}$ Comenius University Bratislava, Faculty of Mathematics, Physics and Informatics, Bratislava, Slovak Republic\\
$^{14}$ Creighton University, Omaha, Nebraska, United States\\
$^{15}$ Department of Physics, Aligarh Muslim University, Aligarh, India\\
$^{16}$ Department of Physics, Pusan National University, Pusan, Republic of Korea\\
$^{17}$ Department of Physics, Sejong University, Seoul, Republic of Korea\\
$^{18}$ Department of Physics, University of California, Berkeley, California, United States\\
$^{19}$ Department of Physics, University of Oslo, Oslo, Norway\\
$^{20}$ Department of Physics and Technology, University of Bergen, Bergen, Norway\\
$^{21}$ Dipartimento di Fisica, Universit\`{a} di Pavia, Pavia, Italy\\
$^{22}$ Dipartimento di Fisica dell'Universit\`{a} and Sezione INFN, Cagliari, Italy\\
$^{23}$ Dipartimento di Fisica dell'Universit\`{a} and Sezione INFN, Trieste, Italy\\
$^{24}$ Dipartimento di Fisica dell'Universit\`{a} and Sezione INFN, Turin, Italy\\
$^{25}$ Dipartimento di Fisica e Astronomia dell'Universit\`{a} and Sezione INFN, Bologna, Italy\\
$^{26}$ Dipartimento di Fisica e Astronomia dell'Universit\`{a} and Sezione INFN, Catania, Italy\\
$^{27}$ Dipartimento di Fisica e Astronomia dell'Universit\`{a} and Sezione INFN, Padova, Italy\\
$^{28}$ Dipartimento di Fisica `E.R.~Caianiello' dell'Universit\`{a} and Gruppo Collegato INFN, Salerno, Italy\\
$^{29}$ Dipartimento DISAT del Politecnico and Sezione INFN, Turin, Italy\\
$^{30}$ Dipartimento di Scienze MIFT, Universit\`{a} di Messina, Messina, Italy\\
$^{31}$ Dipartimento Interateneo di Fisica `M.~Merlin' and Sezione INFN, Bari, Italy\\
$^{32}$ European Organization for Nuclear Research (CERN), Geneva, Switzerland\\
$^{33}$ Faculty of Electrical Engineering, Mechanical Engineering and Naval Architecture, University of Split, Split, Croatia\\
$^{34}$ Faculty of Nuclear Sciences and Physical Engineering, Czech Technical University in Prague, Prague, Czech Republic\\
$^{35}$ Faculty of Physics, Sofia University, Sofia, Bulgaria\\
$^{36}$ Faculty of Science, P.J.~\v{S}af\'{a}rik University, Ko\v{s}ice, Slovak Republic\\
$^{37}$ Faculty of Technology, Environmental and Social Sciences, Bergen, Norway\\
$^{38}$ Frankfurt Institute for Advanced Studies, Johann Wolfgang Goethe-Universit\"{a}t Frankfurt, Frankfurt, Germany\\
$^{39}$ Fudan University, Shanghai, China\\
$^{40}$ Gauhati University, Department of Physics, Guwahati, India\\
$^{41}$ Helmholtz-Institut f\"{u}r Strahlen- und Kernphysik, Rheinische Friedrich-Wilhelms-Universit\"{a}t Bonn, Bonn, Germany\\
$^{42}$ Helsinki Institute of Physics (HIP), Helsinki, Finland\\
$^{43}$ High Energy Physics Group,  Universidad Aut\'{o}noma de Puebla, Puebla, Mexico\\
$^{44}$ Horia Hulubei National Institute of Physics and Nuclear Engineering, Bucharest, Romania\\
$^{45}$ HUN-REN Wigner Research Centre for Physics, Budapest, Hungary\\
$^{46}$ Indian Institute of Technology Bombay (IIT), Mumbai, India\\
$^{47}$ Indian Institute of Technology Indore, Indore, India\\
$^{48}$ INFN, Laboratori Nazionali di Frascati, Frascati, Italy\\
$^{49}$ INFN, Sezione di Bari, Bari, Italy\\
$^{50}$ INFN, Sezione di Bologna, Bologna, Italy\\
$^{51}$ INFN, Sezione di Cagliari, Cagliari, Italy\\
$^{52}$ INFN, Sezione di Catania, Catania, Italy\\
$^{53}$ INFN, Sezione di Padova, Padova, Italy\\
$^{54}$ INFN, Sezione di Pavia, Pavia, Italy\\
$^{55}$ INFN, Sezione di Torino, Turin, Italy\\
$^{56}$ INFN, Sezione di Trieste, Trieste, Italy\\
$^{57}$ Inha University, Incheon, Republic of Korea\\
$^{58}$ Institute for Gravitational and Subatomic Physics (GRASP), Utrecht University/Nikhef, Utrecht, Netherlands\\
$^{59}$ Institute of Experimental Physics, Slovak Academy of Sciences, Ko\v{s}ice, Slovak Republic\\
$^{60}$ Institute of Physics, Homi Bhabha National Institute, Bhubaneswar, India\\
$^{61}$ Institute of Physics of the Czech Academy of Sciences, Prague, Czech Republic\\
$^{62}$ Institute of Space Science (ISS), Bucharest, Romania\\
$^{63}$ Institut f\"{u}r Kernphysik, Johann Wolfgang Goethe-Universit\"{a}t Frankfurt, Frankfurt, Germany\\
$^{64}$ Instituto de Ciencias Nucleares, Universidad Nacional Aut\'{o}noma de M\'{e}xico, Mexico City, Mexico\\
$^{65}$ Instituto de F\'{i}sica, Universidade Federal do Rio Grande do Sul (UFRGS), Porto Alegre, Brazil\\
$^{66}$ Instituto de F\'{\i}sica, Universidad Nacional Aut\'{o}noma de M\'{e}xico, Mexico City, Mexico\\
$^{67}$ iThemba LABS, National Research Foundation, Somerset West, South Africa\\
$^{68}$ Jeonbuk National University, Jeonju, Republic of Korea\\
$^{69}$ Korea Institute of Science and Technology Information, Daejeon, Republic of Korea\\
$^{70}$ Laboratoire de Physique Subatomique et de Cosmologie, Universit\'{e} Grenoble-Alpes, CNRS-IN2P3, Grenoble, France\\
$^{71}$ Lawrence Berkeley National Laboratory, Berkeley, California, United States\\
$^{72}$ Lund University Department of Physics, Division of Particle Physics, Lund, Sweden\\
$^{73}$ Marietta Blau Institute, Vienna, Austria\\
$^{74}$ Nagasaki Institute of Applied Science, Nagasaki, Japan\\
$^{75}$ Nara Women{'}s University (NWU), Nara, Japan\\
$^{76}$ National Centre for Nuclear Research, Warsaw, Poland\\
$^{77}$ National Institute of Science Education and Research, Homi Bhabha National Institute, Jatni, India\\
$^{78}$ National Nuclear Research Center, Baku, Azerbaijan\\
$^{79}$ National Research and Innovation Agency - BRIN, Jakarta, Indonesia\\
$^{80}$ Niels Bohr Institute, University of Copenhagen, Copenhagen, Denmark\\
$^{81}$ Nikhef, National institute for subatomic physics, Amsterdam, Netherlands\\
$^{82}$ Nuclear Physics Group, STFC Daresbury Laboratory, Daresbury, United Kingdom\\
$^{83}$ Nuclear Physics Institute of the Czech Academy of Sciences, Husinec-\v{R}e\v{z}, Czech Republic\\
$^{84}$ Oak Ridge National Laboratory, Oak Ridge, Tennessee, United States\\
$^{85}$ Ohio State University, Columbus, Ohio, United States\\
$^{86}$ Physics department, Faculty of science, University of Zagreb, Zagreb, Croatia\\
$^{87}$ Physics Department, Panjab University, Chandigarh, India\\
$^{88}$ Physics Department, University of Jammu, Jammu, India\\
$^{89}$ Physics Program and International Institute for Sustainability with Knotted Chiral Meta Matter (WPI-SKCM$^{2}$), Hiroshima University, Hiroshima, Japan\\
$^{90}$ Physikalisches Institut, Eberhard-Karls-Universit\"{a}t T\"{u}bingen, T\"{u}bingen, Germany\\
$^{91}$ Physikalisches Institut, Ruprecht-Karls-Universit\"{a}t Heidelberg, Heidelberg, Germany\\
$^{92}$ Physik Department, Technische Universit\"{a}t M\"{u}nchen, Munich, Germany\\
$^{93}$ Politecnico di Bari and Sezione INFN, Bari, Italy\\
$^{94}$ Research Division and ExtreMe Matter Institute EMMI, GSI Helmholtzzentrum f\"ur Schwerionenforschung GmbH, Darmstadt, Germany\\
$^{95}$ Saga University, Saga, Japan\\
$^{96}$ Saha Institute of Nuclear Physics, Homi Bhabha National Institute, Kolkata, India\\
$^{97}$ School of Physics and Astronomy, University of Birmingham, Birmingham, United Kingdom\\
$^{98}$ Secci\'{o}n F\'{\i}sica, Departamento de Ciencias, Pontificia Universidad Cat\'{o}lica del Per\'{u}, Lima, Peru\\
$^{99}$ SUBATECH, IMT Atlantique, Nantes Universit\'{e}, CNRS-IN2P3, Nantes, France\\
$^{100}$ Sungkyunkwan University, Suwon City, Republic of Korea\\
$^{101}$ Suranaree University of Technology, Nakhon Ratchasima, Thailand\\
$^{102}$ Technical University of Ko\v{s}ice, Ko\v{s}ice, Slovak Republic\\
$^{103}$ The Henryk Niewodniczanski Institute of Nuclear Physics, Polish Academy of Sciences, Cracow, Poland\\
$^{104}$ The University of Texas at Austin, Austin, Texas, United States\\
$^{105}$ Universidad Aut\'{o}noma de Sinaloa, Culiac\'{a}n, Mexico\\
$^{106}$ Universidade de S\~{a}o Paulo (USP), S\~{a}o Paulo, Brazil\\
$^{107}$ Universidade Estadual de Campinas (UNICAMP), Campinas, Brazil\\
$^{108}$ Universidade Federal do ABC, Santo Andre, Brazil\\
$^{109}$ Universitatea Nationala de Stiinta si Tehnologie Politehnica Bucuresti, Bucharest, Romania\\
$^{110}$ University of Cape Town, Cape Town, South Africa\\
$^{111}$ University of Derby, Derby, United Kingdom\\
$^{112}$ University of Houston, Houston, Texas, United States\\
$^{113}$ University of Jyv\"{a}skyl\"{a}, Jyv\"{a}skyl\"{a}, Finland\\
$^{114}$ University of Kansas, Lawrence, Kansas, United States\\
$^{115}$ University of Liverpool, Liverpool, United Kingdom\\
$^{116}$ University of Science and Technology of China, Hefei, China\\
$^{117}$ University of Silesia in Katowice, Katowice, Poland\\
$^{118}$ University of South-Eastern Norway, Kongsberg, Norway\\
$^{119}$ University of Tennessee, Knoxville, Tennessee, United States\\
$^{120}$ University of the Witwatersrand, Johannesburg, South Africa\\
$^{121}$ University of Tokyo, Tokyo, Japan\\
$^{122}$ University of Tsukuba, Tsukuba, Japan\\
$^{123}$ Universit\"{a}t M\"{u}nster, Institut f\"{u}r Kernphysik, M\"{u}nster, Germany\\
$^{124}$ Universit\'{e} Clermont Auvergne, CNRS/IN2P3, LPC, Clermont-Ferrand, France\\
$^{125}$ Universit\'{e} de Lyon, CNRS/IN2P3, Institut de Physique des 2 Infinis de Lyon, Lyon, France\\
$^{126}$ Universit\'{e} de Strasbourg, CNRS, IPHC UMR 7178, F-67000 Strasbourg, France, Strasbourg, France\\
$^{127}$ Universit\'{e} Paris-Saclay, Centre d'Etudes de Saclay (CEA), IRFU, D\'{e}partment de Physique Nucl\'{e}aire (DPhN), Saclay, France\\
$^{128}$ Universit\'{e}  Paris-Saclay, CNRS/IN2P3, IJCLab, Orsay, France\\
$^{129}$ Universit\`{a} degli Studi di Foggia, Foggia, Italy\\
$^{130}$ Universit\`{a} del Piemonte Orientale, Vercelli, Italy\\
$^{131}$ Universit\`{a} di Brescia, Brescia, Italy\\
$^{132}$ Variable Energy Cyclotron Centre, Homi Bhabha National Institute, Kolkata, India\\
$^{133}$ Warsaw University of Technology, Warsaw, Poland\\
$^{134}$ Wayne State University, Detroit, Michigan, United States\\
$^{135}$ Yale University, New Haven, Connecticut, United States\\
$^{136}$ Yildiz Technical University, Istanbul, Turkey\\
$^{137}$ Yonsei University, Seoul, Republic of Korea\\
$^{138}$ Affiliated with an institute formerly covered by a cooperation agreement with CERN\\
$^{139}$ Affiliated with an international laboratory covered by a cooperation agreement with CERN.\\

\end{flushleft} 
  
\end{document}